\documentclass{ws-procs9x6}
\def\beq{\begin{equation}}
\def\eeq{\end{equation}}
\def\beqn{\begin{eqnarray}}
\def\eeqn{\end{eqnarray}}

\begin{document}
\title{Effective Lagrangian for Charged Higgs Couplings}
\author{ Tarek Ibrahim$^{a,b}$ and Anastasios  Psinas$^b$}
%\footnote{\uppercase{W}ork partially
%supported by NSF grant PHY-0139967.}}

\address{(a) Department of  Physics, Faculty of Science,
University of Alexandria,\\
 Alexandria, Egypt\\
and\\
(b) Department of Physics, Northeastern University,
Boston, MA 02115-5000, USA}

%ccccccc
\maketitle

\abstracts{
The effective Lagrangian including one loop corrections is deduced for the
couplings of the charged Higgs with quarks and leptons, and with charginos
and neutralinos. The effect of the one loop corrections is found to be 
quite significant in a number of sectors. The effective Lagrangian is then
used to analyze the decay of the charged Higgs into a number of decay
channels. Specifically we consider the decay of $H^+(H^-)$ into the decay modes
  $t\bar b$ ($\bar t b$), $\bar\tau\nu_{\tau}$
 ($\tau \bar \nu_{\tau}$),
and $\chi_i^+\chi_j^0(\chi_i^-\chi_j^0)$ (i=1,2; j=1-4).
The loop corrections to these decay modes are also found to be
quite significant lying in the range 20-30\% in
 significant regions of the parameter space of the SUGRA model.
 The effects of CP phases on the effective Lagrangian and on the
 branching ratios are also analysed and these effects found to be
 important.} 
 
 %%%%%%%%%%%%%%%%%%%%%%%%%%%%%%%
 \section{Introduction}
 Charged Higgs couplings and decays provide an important avenue for
 the exploration of new physics\cite{carena2002}. Recently considerable attention 
 has focussed on one loop corrected effective Lagrangians that enter in 
 the decays $H^+\rightarrow t\bar b$
 ($H^-\rightarrow \bar t b$) and $H^+\rightarrow \bar \tau\nu_{\tau}$ ($H^-\rightarrow \tau\bar\nu_{\tau}$)
 \cite{Carena:1999py,Christova:2002sw,Ibrahim:2003ca,Ibrahim:2003jm,Ibrahim:2003tq}.
 However, in the preceeding works specifically in the works of 
 Refs.\cite{Christova:2002sw,Ibrahim:2003ca,Ibrahim:2003jm,Ibrahim:2003tq},
 the Higgs couplings  with chargino and neutralinos were, not taken into account.
 In this paper we focus on the one loop corrected
 effective Lagrangian including the charged Higgs-chargino-neutralino couplings. 
  The analysis takes into account also the CP phases. 
   The issue of phases is important because of two reasons. First, in 
  MSSM there are a huge number of CP phases that arise in the soft breaking
  sector of the theory. In mSUGRA\cite{msugra} the number of phases is reduced to just 
  two phases, i.e., the phase of the Higgs mixing parameter $\theta_{\mu}$ and
  the phase of the trilinear couplings. Thus mSUGRA 
    is parametrized by the universal scalar mass $m_{0}$, the
 universal gaugino mass $m_{\frac{1}{2}}$, the universal trilinear
coupling $A_{0}$, the ratio of the Higgs vacuum expectation values (VEV's),
 i.e, $ \tan\beta=<H_2>/<H_1> $
 where $H_2$ gives mass to the up quark and
 $H_1$ gives VEV to the down quark  and the lepton. And including the phases
 we have two more parameters, $\theta_{\mu}$ and $\alpha_A$.
 For the non-universal SUGRA the number of parameters increases and so do
 the number of CP phases.
 The inclusion of phases of course draws attention to the
 severe experimental constraints that exist on the electric dipole
 moment (edm) of the electron\cite{eedm}, of the neutron\cite{nedm}
 and of $^{199}Hg$ atom\cite{atomic}.
 However, as is now well known there are a variety of techniques
 available that allow one to suppress the large edms  and bring them
 in conformity with the current
 experiment\cite{na,incancel,olive,chang}.
 Second the CP phases affect a variety of low energy  Thus 
 CP phases affect loop corrections to the Higgs mass\cite{cphiggsmass},
 dark matter\cite{cpdark,gomez} and  a  number of other phenomena
(for a review see  Ref.\cite{Ibrahim:2002ry}).
 The outline of the rest of the paper is as follows: In Sec.2
we compute the loop correction to the
$H^{\pm}\chi^{\mp}\chi^0$ couplings arising from supersymmtric particle
 exchanges and the effects of these corrections on the charged Higgs decay.
 In Sec.4 we give an  analysis of the sizes of radiative corrections.
 It is found that the loop
 correction can be  as  large as 25-30\% in certain parts of
 the parameters space. Conclusions are given in Sec.5.

 \section{Loop Corrections to Charged Higgs Couplings}
We begin with the tree level Lagrangian for $H^{\pm}\chi^{\mp}\chi^0$ interaction 
\beqn
 {{L}}=\xi_{ji}H_2^{1*}\bar \chi_j^0P_L \chi_i^+
 + \xi_{ji}'H_1^2\bar\chi_j^0P_R \chi_i^+ +H.c.
 \label{hchichicouplings}
 \eeqn
 where $H_1^2$ and $H_2^1$ are the charged states of the two Higgs
 iso-doublets in the minimal supersymmetric standard model (MSSM),
 .i.e,
 \beqn
  (H_1)= (H_1^1,H_{1}^2), ~~
 (H_2)= (H_{2}^1,H_2^2) 
\label{a} 
\eeqn 
and $\xi_{ji}$ and $\xi_{ji}'$ are given by
\beqn
\xi_{ji}=-gX_{4j}V_{i1}^*
  -\frac{g}{\sqrt 2} X_{2j} V_{i2}^*
 -\frac{g}{\sqrt 2}\tan\theta_WX_{1j}V_{i2}^*
 \eeqn
 and
 \beqn
  \xi_{ji}'=-gX_{3j}^*U_{i1}
 +\frac{g}{\sqrt 2} X_{2j}^* U_{i2}
 +\frac{g}{\sqrt 2} \tan\theta_WX_{1j}^*U_{i2}
 \eeqn
 where $X$, $U$ and
$V$diagonalize  the neutralino and chargino mass matrices so that
 \beqn
 X^T M_{\chi^0}X= diag (m_{\chi_1^0},
 m_{\chi_2^0},m_{\chi_3^0},m_{\chi_4^0})\nonumber\\
 U^* M_{\chi^+}V^{-1}= diag (m_{\chi_1^+}, m_{\chi_2^+})
 \eeqn where $m_{\chi_i^0}$ (i=1,2,3,4)
 are the eigen values of the neutralino mass matrix
 $M_{\chi^0}$ and $m_{\chi_1^+}, m_{\chi_2^+}$
 are the eigen values of the chargino mass matrix $M_{\chi^+}$.
 The loop corrections produce shifts
 in the couplings of Eq.~(\ref{hchichicouplings})
 and the effective  Lagrangian with loop corrected couplings is
 given by
 \beqn { {L}}_{eff}=(\xi_{ji} + \delta\xi_{ji} ) H_2^{1*}\bar \chi_j^0P_L \chi_i^+
 + \Delta\xi_{ji} H_1^2\bar\chi_j^0 P_L \chi_i^+ \nonumber\\
 + (\xi_{ji}' +\delta \xi_{ji}') H_1^2\bar\chi_j^0P_R \chi_i^+ +
 \Delta \xi_{ji}' H_2^{1*}\bar \chi_j^0P_R \chi_i^+ + H.c.
 \label{hchichicouplings1}
 \eeqn
 As is conventional we calculate the loop correction to the
 $\chi^{\pm}\chi^0H^{\mp}$ using the zero external momentum approximation.
 \begin{figure}
 \hspace*{-0.6in} \centering
 \includegraphics[width=8cm,height=4cm]{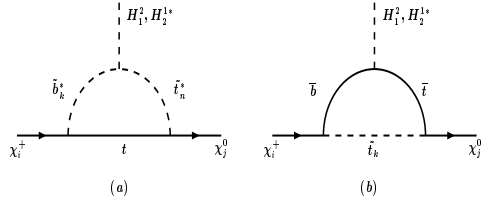}
 \caption{The stop and sbottom exchange
 contributions to the $H^-\chi^+\chi^0$ vertex.} \label{figab}
 \end{figure}
 We note  that the contribution from diagrams which have
 $W-Z-\chi_i^0$ and $W-Z-\chi_i^+$ exchanges in the loop
 vanish due to the absence of $H^+W^-Z$ vertex at tree level. This is a general feature of models
 with two doublets of Higgs\cite{gm}.
 Further, the loops with $H^+W^-H^0_k$ vertices
do not contribute in the zero external momentum approximation
since these vertices are proportional to the external momentum.
Given the fact
 that we ultimately seek to apply the effective
 couplings to the decay
 of the charged Higgs into charginos and neutralinos,
  the mass of the charged
Higgs must be relatively large.
 Consequently,  it is permissible to disregard  diagrams
  which have  $H^{\pm}$ running in the loops due
to the large mass suppression.
   Here as an illustration we give the computation of the loop
   correction corresponding to Fig.(1). 

 For the evluation of $\Delta\xi_{ij}$  for Fig. (1)
 we need $\tilde{b}t\chi^{0}$,  $\tilde{t}t\chi^{0}$ and
 $\tilde{b}\tilde{t}H$ interactions. These are given by
 \begin{equation}
 {{L}}_{\tilde{b}t\chi^{+}}=-g\bar{t}[(U_{l1}D_{b1n}-K_{b}U_{l2}D_{b2n})
 P_{R}-K_{t}V^{*}_{l2}D_{b1n}P_{L}]\tilde{\chi}^+_l\tilde{b_n}+H.c.
 \end{equation}
 \begin{equation}
 { {L}}_{\tilde{t}t\chi^0}=-\sqrt{2}\bar{t}[(\alpha_{tl}D_{t1n}-\gamma_{tl}D_{t2n})P_L+
 (\beta_{tl}D_{t1n}+\alpha^*_{tl}D_{t2n})P_R]\tilde{\chi}^0_l\tilde{t_n}+H.C
 \end{equation}
 \begin{equation}
 { {L}}_{H\tilde{t}\tilde{b}}=H^1_2\tilde{b_k}\tilde{t^*_n}\eta_{kn}+H^2_1\tilde{b^*_k}\tilde{t_n}\eta'_{kn}+H.C
 \end{equation}
 \beqn
 \alpha_{tk} =\frac{g m_tX_{4k}}{2m_W\sin\beta}\nonumber\\
 \beta_{tk}=eQ_tX_{1k}^{'*} +\frac{g}{\cos\theta_W} X_{2k}^{'*}
 (T_{3t}-Q_t\sin^2\theta_W)\nonumber\\
\gamma_{tk}=eQ_t X_{1k}'-\frac{gQ_t\sin^2\theta_W}{\cos\theta_W}
X_{2k}'
 \eeqn
 where $X'$'s are given by
 \beqn
 X'_{1k}=X_{1k}\cos\theta_W +X_{2k}\sin\theta_W\nonumber\\
 X'_{2k}=-X_{1k}\sin\theta_W +X_{2k}\cos\theta_W
\eeqn
and where
 \beqn
 K_{t(b)}=\frac{m_{t(b)}}{\sqrt 2 m_W \sin\beta (\cos\beta)}
 \eeqn
 \begin{figure}
 \hspace*{-0.6in}
 \centering
 \includegraphics[width=8cm,height=4cm]{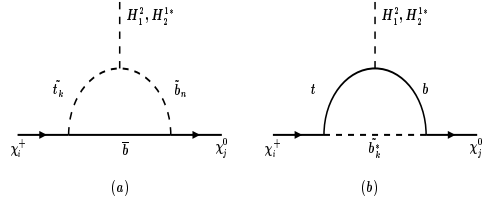}
 \caption{Another set of  diagrams exhibiting  stop and sbottom exchange
 contributions to the $H^-\chi^+\chi^0$ vertex. }
 \label{figcd}
 \end{figure}
 Finally, $\eta_{ij}$ is defined by
 \beqn\label{f}
 \eta_{ij}= \frac{gm_t}{\sqrt 2 m_W \sin\beta} m_0A_tD_{b1i} D_{t2j}^*
 +\frac{gm_b}{\sqrt 2 m_W \cos\beta}\mu D_{b2i} D_{t1j}^*\nonumber\\
 +\frac{gm_bm_t}{\sqrt 2 m_W \sin\beta} D_{b2i} D_{t2j}^*
  +\frac{gm_t^2}{\sqrt 2 m_W \sin\beta} D_{b1i} D_{t1j}^*
 -\frac{g}{\sqrt 2} m_W \sin\beta D_{b1i} D_{t1j}^*
 \eeqn
 and $\eta_{ij}'$ is defined by
 \beqn\label{e}
 \eta_{ji}'= \frac{gm_b}{\sqrt 2 m_W \cos\beta} m_0A_bD_{b2j}^* D_{t1i}
 +\frac{gm_t}{\sqrt 2 m_W \sin\beta} \mu D_{b1j}^* D_{t2i}\nonumber\\
 +\frac{gm_bm_t}{\sqrt 2 m_W \cos\beta}  D_{b2j}^* D_{t2i}
 +\frac{gm_b^2}{\sqrt 2 m_W \cos\beta} D_{b1j}^* D_{t1i}
 -\frac{g}{\sqrt 2} m_W \cos\beta D_{b1j}^* D_{t1i}
 \eeqn
 where $D_{bij}$ is the matrix that
  diagonalizes the b squark $mass^2$ matrix so that
 \beqn \tilde b_L=\sum_{i=1}^{2} D_{b1i} \tilde b_i,~~~~~
 \tilde b_R=\sum_{i=1}^{2} D_{b2i} \tilde b_i
 \eeqn
  where $\tilde b_i$ are the b squark mass eigen states.
 In a similar fashion  $D_{tij}$ diagonalizes the t squark
 $mass^2$ matrix so that
 \beqn \tilde t_L=\sum_{i=1}^{2} D_{t1i}
 \tilde t_i,~~~~~
 \tilde t_R=\sum_{i=1}^{2} D_{t2i} \tilde t_i
 \eeqn
 where $\tilde t_i$ are the t squark mass eigen states.
 Using the above one finds for Fig. (1)
 the result\cite{inp}
 \beqn
 \Delta\xi^{(1a)}_{ji}=-\sum_{k=1}^2\sum_{n=1}^2\sqrt{2}gK_tV^*_{i2}D_{b1k}\eta'_{kn}
 (\beta^*_{tj}D^*_{t1n}+\alpha_{tj}D^*_{t2n})\nonumber\\
 (\frac{m_t}{16\pi^2})
 f(m^2_t,{m}^2_{\tilde b_k},{m}^2_{\tilde t_n})
 \eeqn
 where the form factor $f(m^2,m_i^2,m_j^2)$ is defined for $i\neq j$ so that
 \beqn\label{h} f(m^2,m_i^2,m_j^2)
 = \frac {1}{(m^2-m_i^2) (m^2-m_j^2)(m_j^2-m_i^2)}\nonumber\\
 (m_j^2 m^2 ln\frac{m_j^2}{m^2}
  +m^2 m_i^2ln\frac{m^2}{m_i^2} +m_i^2 m_j^2 ln\frac{m_i^2}{m_j^2})
 \eeqn
 and for the case $i= j$ it is given by
 \beqn\label{i}
 f(m^2,m_i^2,m_i^2) =\frac {1}{(m_i^2-m^2)^2} (m^2 ln\frac{m_i^2}{m^2}
 + (m^2-m_i^2))
\eeqn

\section{Charged Higgs Decays Including Loop Effects}
Using the above analysis the  effective Lagrangian 
for $H^{\pm}{\chi^{\mp}} \chi^0$ with loop corrections 
 may be written as follows
 \begin{equation}
 {{L}}_{eff}=H^-\overline{\chi^0_j}(\alpha^S_{ji}+\gamma_{5}\alpha^P_{ji})\chi^+_i+
 H.c
\end{equation}
where
\begin{equation}
\alpha^{S}_{ji}=\frac{1}{2}(\xi'_{ji}+\delta\xi'_{ji})\sin{\beta}+
 \frac{1}{2}\Delta\xi'_{ji}\cos{\beta}+\frac{1}{2}(\xi_{ji}+
\delta\xi_{ji})\cos{\beta}+\frac{1}{2}\Delta\xi_{ji}\sin{\beta}
 \label{scoupling}
\end{equation}
and where
\begin{equation}
\alpha^{P}_{ji}=\frac{1}{2}(\xi'_{ji}+\delta\xi'_{ji})\sin{\beta}
 +\frac{1}{2}\Delta\xi'_{ji}\cos{\beta}-
\frac{1}{2}(\xi_{ji}+\delta\xi_{ji})\cos{\beta}-
\frac{1}{2}\Delta\xi_{ji}\sin{\beta} \label{pcoupling}
 \end{equation}
The effective couplings contain dependence on CP phases and thus
the branching ratios will be sensitive to the CP phases. 
Such dependence arises via  the diagonalizing matrices U and V
 from the chargino sector and via the matrix X in the neutralino sector.
The decay width of $H^-$ into $\chi^0_j\chi^-_i$ (j=1,2; i=1,2,3,4)
is given by 
\beqn
\Gamma_{ji}(H^{-}\rightarrow\chi^0_j\chi^-_i)=\frac{1}{4\pi
 M^3_{H^-}}
 \sqrt{[(m^2_{\chi^0_j}+m^2_{\chi^{+}_i}-M^2_{H^{-}})^2
 -4m^2_{\chi^{+}_i}m^2_{\chi^0_j}]}\nonumber\\
 ([\frac{1}{2}((|\alpha^{S}_{ji}|)^2+(|\alpha^{P}_{ji}|)^2)
 (M^2_{H^{-}}-m^2_{\chi^{+}_i}-m^2_{\chi^{0}_j})\nonumber\\
 -\frac{1}{2}((|\alpha^{S}_{ji}|)^2-(|\alpha^{P}_{ji}|)^2)
 (2m_{\chi^{+}_i}m_{\chi^{0}_j})])
 \label{branching}
 \eeqn
 Here the CP phase dependence will arise from the fact that the 
 chargino and neutralino masses are sensitive to CP phases and
 also from the dependence of the effective couplings on the 
 phases.

\begin{figure}
% \hspace*{-0.6in}
 \centering
 \includegraphics[width=10cm,height=6cm]{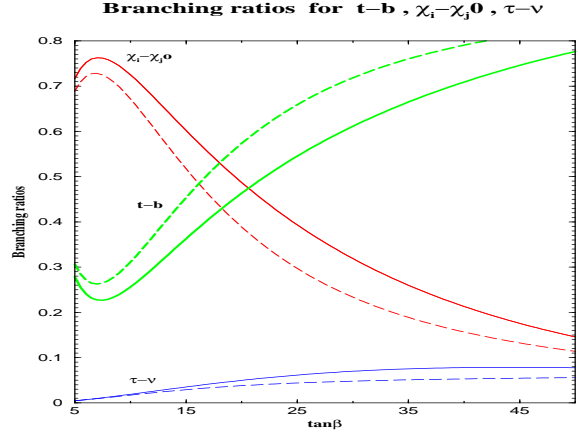}
 \caption{
 Plot of branching ratios for the decay of $H^{\pm}$ as a function of 
$\tan\beta$. The  parameters are 
$m_A=800$, $m_0=400$, $m_{\frac{1}{2}}=140$, $A_0$=3, 
 $\xi_1=0$,  $\xi_2=0$,  $\xi_3=0$,  $\theta_{\mu}=0$,
 $\alpha_{A_0}=0$. The long dashed lines are the branching
ratios at the tree level while the solid lines include the 
loop correction. The curves labelled $\chi_i^-\chi_j^0$
stand for sum of branching ratios into all allowed 
$\chi_i^-\chi_j^0$ modes. All masses are in unit of GeV and 
all angles in unit of radian. From Ref.(7).}                       
 \label{fig5b}
 \end{figure}

\begin{figure}
% \hspace*{-0.6in}
 \centering
 \includegraphics[width=10cm,height=6cm]{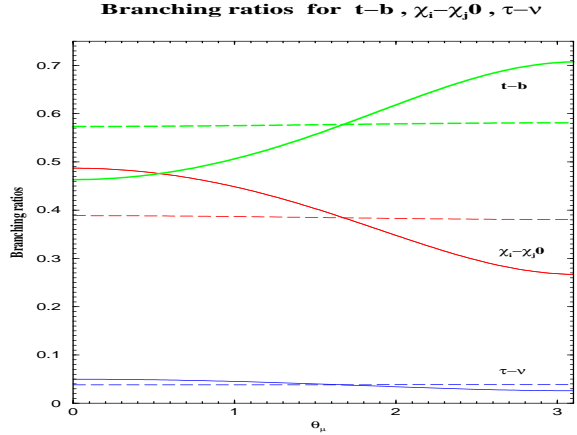}
 \caption{
Plot of branching ratios for the decay of $H^{\pm}$ as a function of 
$\theta_{\mu}$  as a function of $\alpha_{A_0}$ in (b).
 The  parameters are 
$m_A=800$, $m_0=400$, $m_{\frac{1}{2}}=140$, $A_0$=3, $\tan\beta=20$,  
 $\xi_1=0$,  $\xi_2=0$,  $\xi_3=0$,  $\theta_{\mu}=0$,
 $\alpha_{A_0}=0$.  
 The long dashed lines are the branching
ratios at the tree level while the solid lines include the 
loop correction. All masses are in unit of GeV and all angles in 
unit of radian. From Ref.(7). }                                   
\label{figmuedm} 
 \end{figure}

\section{Sizes of Loop Corrections}
The theoretical analysis of effective Lagrangian obtained here 
including loop corrections is quite general. However, the parameter
space of the general MSSM is quite large and thus we investigate
the sizes of the effects in more constrained parameter space.
This more constrained parameter space is provided by the extended
SUGRA model. Thus we assume that the parameter space
of the model to consist of 
 $m_A$ (mass of the  CP odd Higgs boson),
 $\tan\beta$, complex trilinear coupling
$A_0$,   $SU(3)$, $SU(2)$ and $U(1)_Y$ gaugino masses
 $\tilde m_i =m_{\frac{1}{2}} e^{i\xi_i}$ (i=1,2,3) and $\theta_{\mu}$, where
 $\theta_{\mu}$ is the phase of $\mu$.  In the analysis the soft parameters
 are evolved from the grand unification scale to the electroweak scale.
 Further, as is usually the case the $\mu$ parameter is determined by the
 constraint of electroweak symmetry breaking while the phase of $\mu$, i.e,,
 $\theta_{\mu}$ remains an arbitrary parameter. It should be noted that 
 while there are several phases in the analysis not all of them are 
 independent\cite{inmssm}.

The main modes of decay of the charged Higgs consist of final states
which include top-bottom, chargino-neutralino ,and tau-neutrino .
In Fig.~\ref{fig5b} we give a plot of the branching ratios
of $H^-$ to $\bar t b$, $\tau^- \bar \nu_{\tau}$ and
 $\chi_i^-\chi_j^0$ as a function of $\tan\beta$. For comparison the
 tree level branching ratio and the loop corrected branching ratios are
 plotted. The analysis of Fig.~\ref{fig5b} shows that the loop corrections
 can be substantial and can reach as much as 20\% or more. 
 In  Fig.~\ref{figmuedm} an analysis of the branching ratios for
 top-bottom, chargino-neutralino ,and tau-neutrino  as a function 
 of $\theta_{\mu}$ is given with and without loop corrections.
 The plots are given as a function of the $\mu$ phase.
 One finds that while the tree level analysis is independent of 
 the phases the loop corrected branching ratios show a rather large
 dependence. The analysis illustrates both the importance of the 
 loop corrections as well as dependence on the phases. 
 Also of interest is the phenomenon of trileptonic signal arising from
 the decay of $H^{\pm}$. This arises when $H^{\pm}$
  decays into a $\chi^{\pm}_1\chi_2^0$ with
 subsequent decays of $\chi^{-}_1$ and $\chi_2^0$ can provide a
trileptonic mode. Thus, e.g., 
 $H^{-}\rightarrow \chi^{-}_1 \chi_2^0$
  $\rightarrow$ ${\it l_1^{-}{\it l_2^+} {\it l}_2^-}$.
 Such a signal is well known in the
context of the decay of the W boson. For off shell decays
it was discussed  in Ref.\cite{Nath:sw}. (For a more recent analysis see Ref.\cite{sugratrilep}).
For the Higgs decay here, the signal can appear for on shell decays since
the mass of the Higgs is expected to be
large enough for such a decay to occur on shell. The analysis presented
here shows that the effect of the loop corrections and of the 
CP phase $\theta_{\mu}$ on these signals can be substantial. 
The effect of other CP phases, e.g., $\xi_3$, $\alpha_A$ on the 
couplings and on the branching ratios can also be significant\cite{inp}. .

\section{Conclusion}
In this paper we have discussed the effective Lagrangian at the one
loop level for the charged higgs-chargino-neutralino interactions.
This analysis augments the previous such analyses for the 
$H^+\bar t b$ and  $H^+\tau\bar \nu$ type couplings. The analysis
presented here includes the dependence of the couplings on the
supersymmetric CP phases. 
One of the interesting results of the analysis is the phenomenon
 that the supersymmetric loop corrections are generally quite 
 substantial ,as much as 20-30\% in significant
 regions of the parameter space of the theory.
We also analysed in this paper the effects of the loop
corrections on the decays of the charged Higgs. Specifically
the analysis of the decay $H^-\rightarrow \chi_1^-\chi_2^0
\rightarrow l_1^-l_2^-l_2^+$, which is the well known 
trileptonic signal,  shows that the loop effects 
here are significant reaching as high as 20-30\%. 
Finally it is found that the loop corrected couplings are 
quite sensitive to CP phases. The effective Lagrangian
presented here should be of considerable interest in the analysis
of charged Higgs decays and for the search for supersymmetry.

\noindent
{\bf Acknowledgments}\\
This research was also supported in part by NSF grant PHY-0139967.\\

%%%%%%%%%%%%%%%%%%%%%%%%%%%%%%%%%%%%%%%%%%%%%%%%%%%%%%%%

\end{document}